\documentclass[aps,prb,twocolumn,groupedaddress]{revtex4}
\usepackage{graphicx}

\begin{document}


\title{Quantum shape effects on Zeeman splittings in semiconductor
  nanostructures}


\author{Pochung Chen}
\affiliation{Department of Physics, National Tsing-Hua University,
  Hsinchu 100, Taiwan}


\date{\today}

\begin{abstract}
  We develop a general method to calculate Zeeman splittings of
  electrons and holes in semiconductor nanostructures within the
  tight-binding framework. The calculation is carried out in the 
  electron-hole picture and is extensible to the excitonic calculation
  by including the electron-hole Coulomb interaction. The method
  is suitable for the investigation of quantum shape effects and the
  anisotropy of the g-factors. Numerical results for CdSe and CdTe
  nanostructures are presented.
\end{abstract}


\maketitle

\section{Introduction}

Controllability of spins in semiconductor nanostructures, or quantum
dots, has become one of the important subject to be investigated in
recent years due to the novel field of spintronics\cite{zutic:323} and quantum
information processing.\cite{Nielsen00} Manipulation of the spin depends crucially on
fundamental spin properties such as the effective Land{\'e} g-factors.\cite{kosaka2001}
On the other hand it is known that quantum confinement gives rise to novel electronic
and spin properties and the shape control of nanostructures has
become possible recently.\cite{xiaogangpeng00,manna00} 
It is thus imperative to understand what kind of 
quantum effects on magneto-optical properties can be induced by changing
the size and the shape of nanostructures. 

The size and shape effects on g-factors in nanostructures have been
investigated within the $\mathbf{k} \cdot \mathbf{p}$ framework
\cite{kotlyar:085310,prado:201310,rodina:155312} as well as 
the tight-binding framework.\cite{chen:045311,schrier:235301} 
Within the $\mathbf{k} \cdot \mathbf{p}$ framework the quantum
confinement and shape effects are taken into account by imposing
various boundary conditions.  However, for the smaller size
nanostructures the atomistic effects may become more relevant, and it
is more difficulty to capture those effects with $\mathbf{k} \cdot
\mathbf{p}$ method.  For example, the mixing between heavy-hole band,
light-hole band, and the spin-orbit split band become stronger as the
size of the nanostructures goes down. The symmetry can be reduced,
depending on the particular shape of the nanostructure. On the other
hand it has been proven that the tight-binding approach can capture the
atomic nature and surfaces effects. It does not need to assume the
nanostructures possessing any symmetry, and has been used to 
calculate the g-factors of CdSe nanocrystals
with reasonable success.\cite{chen:045311,schrier:235301}

In this work we develop a general method which is suitable for the
investigation of quantum shape effects on both electron and hole g-factors.
We emphasize that we are working in the electron-hole picture instead of
the conduction-valence band electron picture. 
This is one of the major difference in comparison with some of the preceding
work using tight-binding method. \cite{chen:045311,schrier:235301}
The electron-hole picture is more relevant to
the real experiment\cite{gupta98} and the method can be extended to 
calculate the exciton Zeeman splitting in a straightforward manner. 
The magnetic field is assumed to be in the weak magnetic regime, in
which the cyclotron length $l_B=\sqrt{\hbar c/e|B|}$ is larger than
the size of the nanostructure. The quantum shape effects are
investigated in detail. 
In the following we will first describe how to setup the Hamiltonian
in electron-hole picture including the effect of vector potential associated
with the external magnetic field.
We then discuss in detail how to extract Zeeman splittings based
on this Hamiltonian. Numerical results for CdSe and CdTe nanostructures
will be presented and discussed. In the end we will discuss the effect
of Coulomb interaction which is omitted in this work.

\section{Method}
The first step in the whole analysis is to obtain the single particle
eigenfunctions and eigenenergies based on a semi-empirical
tight-binding method, resulting in a Hartree-Fock ground state. The
spin-orbit interaction is included in the tight-binding Hamiltonian.
The dangling bonds are truncated. An electron-hole transformation over
the ground state is employed to obtain the Hamiltonians in the
electron-hole picture.\cite{Haken83} The second quantization field
operator is written as
\begin{eqnarray}
  \psi(x)
  &=&
  \sum_{l}a_{l+}\left(\begin{array}{c} \phi_{l}(x) \\ 0\end{array}\right)
  +\sum_{l}a_{l-}\left(\begin{array}{c} 0\\\phi_{l}(x)\end{array}\right) \\
  &=& \nonumber
  \sum_{k_{c}}a_{k_{c}}
  \left(\begin{array}{c} \phi_{k_{c}+}(x) \\ \phi_{k_{c}-}(x)\end{array}\right)
  +\sum_{k_{v}}a_{k_{v}}
  \left(\begin{array}{c}\phi_{k_{v}+}(x)\\ \phi_{k_{v}-}(x)\end{array}\right),
\end{eqnarray}
where $a_{l\sigma}$ is the annihilation operator of local orbitals
with collective site-orbital index $l$ and spin index $\sigma$, and
$a_{k_c} (a_{k_v})$ is the annihilation operator of the conduction
(valence) band electrons. The field operator expanded in local basis
gives rise to the typical single-particle tight-binding Hamiltonian
\begin{equation}
  \label{eq:H_tb}
  H_{\mathrm{T.B.}}=
  \sum_{l\sigma,l^\prime \sigma^\prime} t_{l\sigma,l^\prime \sigma^\prime}  
  a^\dagger_{l\sigma,l^\prime} a_{l^\prime \sigma^\prime},
\end{equation} 
where the summation is restricted to the on-site and nearest-neighbor
pairs in this work. We use the well-known semi-empirical values for
the tight-binding parameters.\cite{leung98} By diagonalizing this
Hamiltonian the eigenenergies $E_{k}$, the relation between
$a_{l\sigma}$ and $a_{k_i}$, as well the relation between
$\phi_{l}(x)$ and $\phi_{k_i}(x)$ can be derived. To transform the
Hamiltonian to the electron-hole picture we define the hole operator
to be $h_{k_v}\equiv a^\dagger_{k_v}$. In this picture the free
Hamiltonian without external magnetic field becomes
\begin{equation}
  H_0=\sum_{k_c} E_{k_c} 
  a^\dagger_{k_c} a_{k_c}-\sum_{k_v} E_{k_v} h^\dagger_{k_v} h_{k_v}.
\end{equation}
Note that $E_{k_c}>0$, $E_{k_v}<0$, and we have dropped the overall
energy constant of the Hartree-Fock ground state.

The external magnetic field will introduce two additional terms to the
Hamiltonian, giving rise to Zeeman splittings.  The first contribution
comes from the Zeeman Hamiltonian:
\begin{equation}
  H_{Zeeman} =
  \int d^{3}x\sum_{\sigma\sigma^{\prime}}
  \psi_{\sigma}^{\dagger}(x)
  g_0 \mu_B \left(\vec{B}\cdot\vec{S}\right)_{\sigma\sigma^{\prime}}
  \psi_{\sigma^{\prime}}(x),
\end{equation}
where $g_0=2.0$ is the free electron g-factor, $\mu_B$ is the Bohr magneton,
$\vec{S}=\frac{1}{2}\vec{\sigma}$, and $\vec{B}$ represents the
external magnetic field. It can be shown that in the electron-hole
picture the Zeeman Hamiltonian becomes
\begin{eqnarray}
  \label{eq:H_Zeeman}
  H_{Zeeman} 
  & = & 
  \sum_{k_{c}k_{c}^{\prime}} g_0 \mu_B e_{k_{c}}^{\dagger}e_{k_{c}^{\prime}}
  \sum_{l}\sum_{\sigma\sigma^{\prime}} 
  f_{k_{c}\sigma}^{l*}
  \vec{B}\cdot\vec{S}_{\sigma\sigma^{\prime}}
  f_{k_{c}^{\prime}\sigma^{\prime}}^{l}\\
  & - & \nonumber
  \sum_{k_{v}k_{v}^{\prime}} g_0 \mu_B h_{k_{v}^{\prime}}^{\dagger}h_{k_{v}}
  \sum_{l}\sum_{\sigma\sigma^{\prime}}
  f_{k_{v}\sigma}^{l*}
  \vec{B}\cdot\vec{S}_{\sigma\sigma^{\prime}}
  f_{k_{v}^{\prime}\sigma^{\prime}}^{l},
\end{eqnarray}
where we have defined $f_{k_{i}\sigma}^{l}=\int
d^{3}x\phi_{l}^{*}(x)\phi_{k_{i}\sigma}(x)$. Note that one should be
careful about the indices of the hole operators.  Within the
tight-binding framework, $f_{k_i\sigma}^l$ simply means the
coefficients of the eigenfunctions in terms of local orbitals.  The
second contribution comes from the vector potential associated with
the external magnetic field. It is known that the effect of vector
potential $\vec{A}$ can be incorporated into the single-particle
tight-binding Hamiltonian\cite{boykin01} by modifying the Hamiltonian
to be
\begin{equation}
  H_A=\sum_{l\sigma,l^\prime \sigma^\prime} t_{l\sigma,l^\prime \sigma^\prime}  
  e^{-iX_{ll^\prime}} a^\dagger_{l\sigma,l^\prime} a_{l^\prime \sigma^\prime},
\end{equation} 
where
\begin{equation}
  X_{ll^\prime}(\vec{B}) \equiv \frac{e}{\hbar}\int_{\vec{R}_{l^{\prime}}}^{\vec{R}_{l}} 
  \vec{A}(\vec{B})\cdot d\vec{r}.
\end{equation} 
Transforming the Hamiltonian into electron-hole picture, one finds
that
\begin{eqnarray}
  \label{eq:H_A}
  H_{A} 
  & = & 
  H_0 + E(\vec{B}) \\
  & + & \nonumber
  \sum_{k_{c}k_{c}^{\prime}}e_{k_{c}}^{\dagger}e_{k_{c}^{\prime}}
  \sum_{l\sigma l^{\prime}\sigma^{\prime}}t_{l\sigma l^{\prime}\sigma^{\prime}}
  \left( e^{-iX_{ll^{\prime}}} -1 \right)
  f_{k_{c}\sigma}^{l*}f_{k_{c}^{\prime}\sigma^{\prime}}^{l^{\prime}}\\
  & - & \nonumber
  \sum_{k_{v}k_{v}^{\prime}}h_{k_{v}^{\prime}}^{\dagger} h_{k_{v}}
  \sum_{l\sigma l^{\prime}\sigma^{\prime}}t_{l\sigma l^{\prime}\sigma^{\prime}}
  \left( e^{-iX_{ll^{\prime}}} -1 \right)
  f_{k_{v}\sigma}^{l*}f_{k_{v}^{\prime}\sigma^{\prime}}^{l^{\prime}},  
\end{eqnarray}
where
\begin{equation}
  E(\vec{B})=
  \sum_{k_{v}}
  \sum_{l\sigma l^{\prime}\sigma^{\prime}}t_{l\sigma l^{\prime}\sigma^{\prime}}
  \left(e^{-iX_{ll^{\prime}}}-1\right)  
  f_{k_{v}\sigma}^{l*}f_{k_{v}\sigma^{\prime}}^{l^{\prime}}.
\end{equation}
represents a small overall energy shift of the ground state due to the
vector potential. It should be noted that during the transformation we
have dropped the terms which do not separately conserve the electron
and hole numbers.\cite{Haken83} Note also that there is no magnetic
field induced coupling between electrons and holes within this
approximation.

In this work the magnitude of the magnetic field is restricted to less
than 15 Tesla. For an uniform field $|\vec{B}|<15 $T and size of
nanostructure less than 100 \AA\ in average diameter, it is easy to
estimate that $X(\vec{B})$ is at the order of $10^{-2}$ to $10^{-3}$,
which means $e^X-1\approx X$.  As a result each term in $H_A$ is
roughly linear in $\vec{B}$ for the parameter range we are interested
in.  To summarize, the two-particle electron-hole Hamiltonian
including the effect of external magnetic field is represented by
\begin{equation}
  \label{eq:H_eh}
  H_{\mathrm{eh}}=H_{\mathrm{Zeeman}}+H_{\mathrm{A}},
\end{equation}
where $H_{Zeeman}$ is defined by Eq. \ref{eq:H_Zeeman} and $H_A$ is
defined by Eq. \ref{eq:H_A}.


\section{Zeeman splittings}
In this section we will briefly discuss how to extract Zeeman
splittings based on Eq. \ref{eq:H_eh}, starting from the single
particle Hamiltonian Eq. \ref{eq:H_tb}. The first step is to find the
single particle eigenenergies and eigenfunctions of Eq. \ref{eq:H_tb}.
Once the single particle eigenenergies and eigenstates are known, the
Hamiltonian in electron-hole picture,
$H_{\mathrm{eh}}=H_{\mathrm{Zeeman}}+H_{\mathrm{A}}$, can be readily
constructed.  In this work we have used exact diagonalization for
nanostructures with less than $777$ atoms. It will be shown later that
the full spectrum of the single particle Hamiltonian is not necessary
if we are only interested in the Zeeman splitting of states within
some energy window. Hence for larger nanostructures one may use
Lanczos or other methods to evaluate the single-particle states
within some energy window, to reduce the computational demand.

It is evident from the Hamiltonian, $H_{\mathrm{eh}}$, that the
external magnetic field induces coupling between all zero-field
eigenstates. In principle in order to evaluate the eigenenergies at
non-zero magnetic field one has to diagonalize the whole Hamiltonian,
including all the electron levels or hole levels. However in
nanostructures, quantum confinement effects give rise to a discrete
density of states, lifting the degeneracies. In particular for the
valence band, the heavy-hole light-hole degeneracy is usually lifted.
As a result, typically the only degeneracy left is the Kramers'
degeneracy. For a given a Kramers' doublet, the external magnetic
field will remove the Kramers' degeneracy and introduce a Zeeman
splitting for the doublet.  For small external magnetic field the
coupling to other doublets can be neglected compared to the coupling
within the doublets, since there is a finite energy seperation between
different doublets. It is thus convenient to define an {\em intrinsic}
Zeeman splitting, which is determined by restricting the Hamiltonian
to the Hilbert space spanned by one particular Kramers' doublet.

As stated near the end of last section, for the magnetic field we are
interested in, an intrinsic Zeeman splitting is nearly linear in B. We
define the {\em intrinsic} g-factor as the ratio between the intrinsic
Zeeman splitting and the magnitude of the magnetic field (up to a
sign).  When the intrinsic Zeeman splitting is smaller than the
inter-level spacing the effects from nearby doublets are small, and
the intrinsic Zeeman splitting is close to the true Zeeman splitting.
However, when the intrinsic Zeeman splitting becomes comparable to the
inter-level spacing the effects from nearby doublets become important.
The intrinsic Zeeman splitting may substantially deviate from the true
Zeeman splitting. In this case, it becomes necessary to include nearby
doublets in order to find the true Zeeman splitting.

We have numerically verified that typically less than 5 nearby
doublets are needed to be included.  Inclusion of further states does
not change the Zeeman splitting. Since we are interested in the states
nearby the band edge, we only need to find the single-particle levels
near the band edge, eliminating the need to find the whole spectrum
and eigenfunctions.  When more than one Kramers' doublet is included
in the calculation, the Zeeman splitting may become strongly
non-linear in B, or level (anti-)crossing might occur. In those cases
the g-factor becomes ill defined, but one can still define an
intrinsic g-factor based on the intrinsic Zeeman splitting.

The Zeeman splitting can be consider to be always positive, while we
usually associate a sign to the g-factor. It is thus important to
determine the sign of the g-factor in a consistent way.  Due to the
spin-orbit interaction and quantum confinement, there is a difference
in spin content between the free carriers and the carriers in
nanostructures. In this work we use the intrinsic Zeeman splitting to
determine the sign of the (intrinsic) g-factor. Basically one has to
find the correct zeroth order eigenstates within the Kramers'
degenerate space, and use the spectrum weight of zeroth order
eigenstates to connect to the real spin.

As an example to illustrate the general numerical procedure to assign
the sign of the g-factor, consider the case in which the external
magnetic is pointing along in the z-direction. The first step is to
find the eigenfunctions of $H_{\mathrm{eh}}$ when a very small
magnetic field is added. One then calculates the spectrum weight of
the two eigenfunctions $|\eta\rangle,|\eta^\prime\rangle$ and connects
them to the real spin using their major spin component, for example,
$|\eta\rangle\rightarrow |s+\rangle$ and
$|\eta^\prime\rangle\rightarrow |s-\rangle$. Now if a zeroth order
eigenfunction $|\eta\rangle$ moves toward higher energy as one
increases the external magnetic field, the g-factor is assigned to be
positive. In this analysis one must ensure that the spin quantization
axis aligned with the external magnetic field. As a result an unitary
transformation on the eigenfunction might be needed when the magnetic
field is not along the z-direction, the typical spin quantization
axis. This assignment may also become impractical when it becomes
difficult to identify the major spin component of zeroth order
eigenstates. In the results described below, we have found it
difficult to identify the major spin component for some of the hole
states.

\section{Results}

\begin{figure}
  \includegraphics[scale=0.35,angle=0]{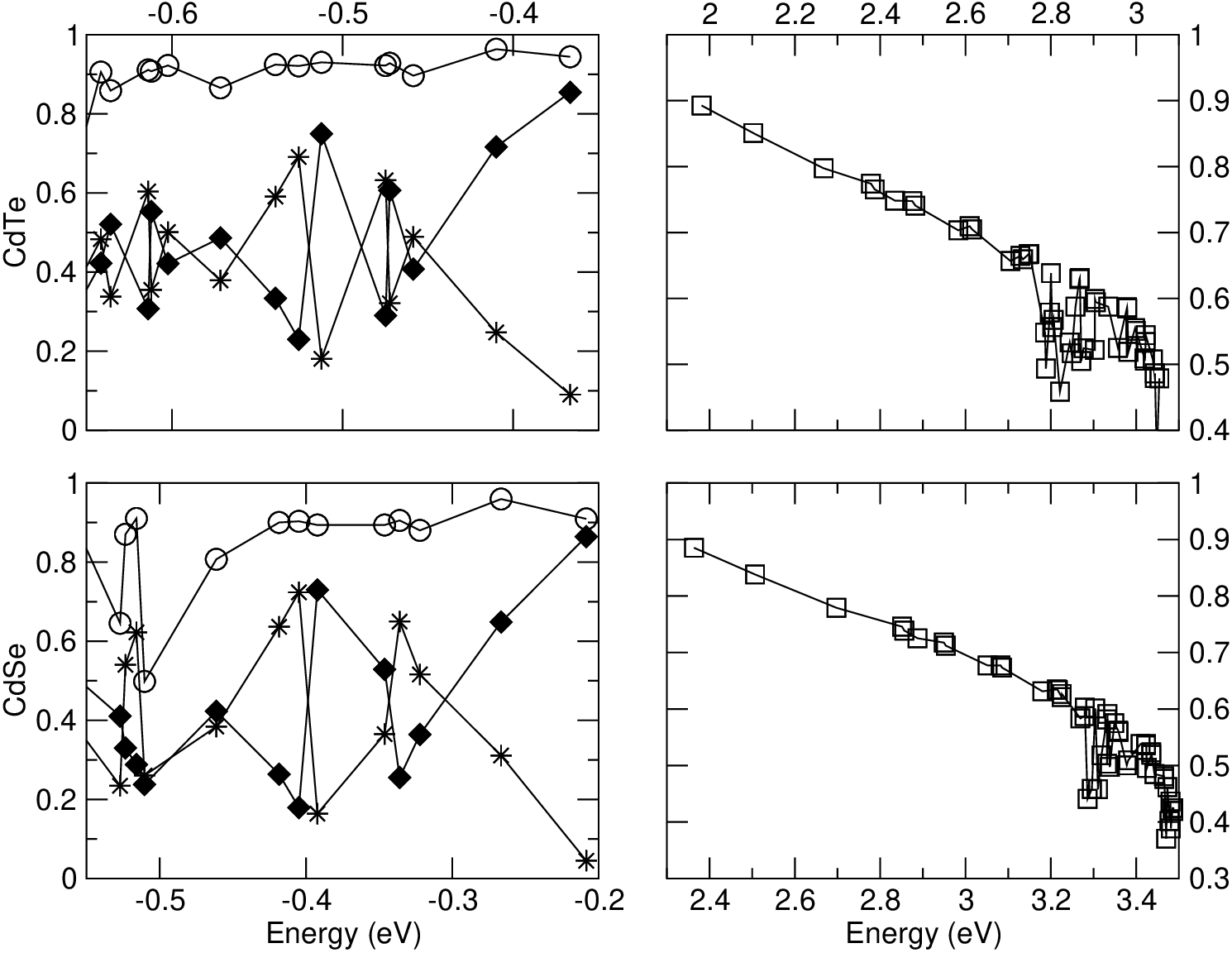}
  \caption{Spectra weight of single particle
    eigenstates.\label{Fig:SpectraWeight}}
\end{figure}

In the bulk limit, the conduction band is $S=\frac{1}{2}$ like and the
valence band is $J=\frac{3}{2}$ like. We calculate the spectra weight
of zero-field single particle eigenstates in terms of local orbitals.
In all calculations we adopt an coordinate system in which the z-axis
is parallel to the c-axis of the wurtzite structure.  We find that for
the conduction band, around $90$\% of the spectra weight of the band
edge state is in $S=\frac{1}{2}$, which increases as the size of
nanostructures increases. However for the same nanostructure, the
$S=\frac{1}{2}$ component decreases as we move toward higher energy
condunction electron states.  For the valence band, we find that the
band is $J=\frac{3}{2}$ like within a fair large energy window near
the valence band edge , containing around 90\% of the spectra weight.
However the mixing between heavy-hole and light-hole is strong and the
mixing is sensative to the size and shape of the nanostructure. It is
thus improper to assign the label of heavy-hole or light-hole to the
holes states in those nanostructures. In Fig.\ref{Fig:SpectraWeight}
we plot the spectra weight of zero-field eigenstates near the band
edges for the 777 atoms CdTe and CdSe nanostructures. Nanostructures
containing different number of atoms show qualitatively similar
behavior. The main features to be noticed are the reduction of
$S=\frac{1}{2}$ ($S=\frac{3}{2}$) components for the electron (hole)
states as we move away from the conduction (valence) band edge, and
the strong heavy-hole light-hole mixing.

\begin{figure}
  \includegraphics[scale=0.35,angle=0]{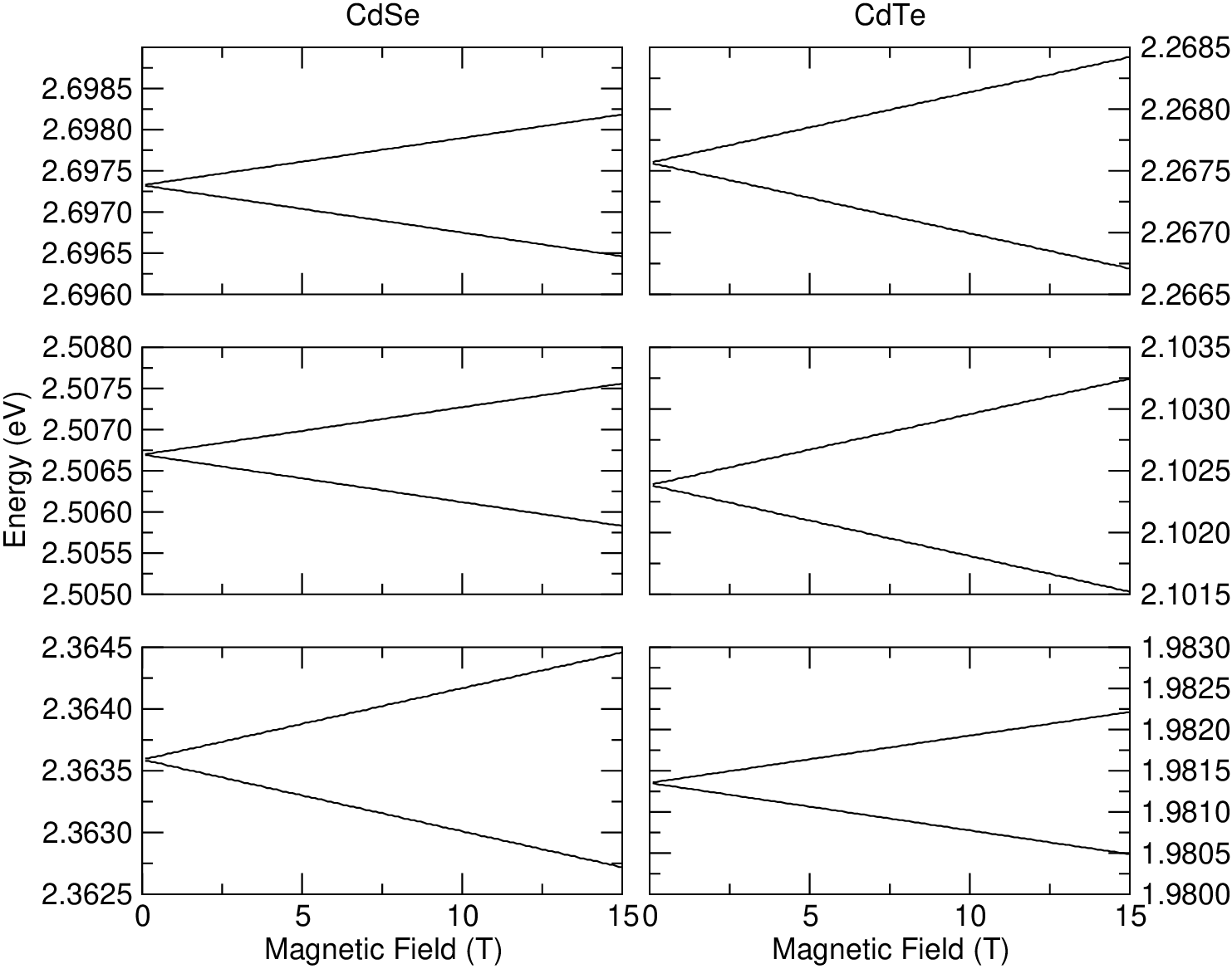}
  \caption{Electron Zeeman splitting for 777 atoms CdSe and CdTe nanostructures.
    \label{Fig:Zeeman777}}
\end{figure}

\begin{figure}
  \includegraphics[scale=0.35,angle=0]{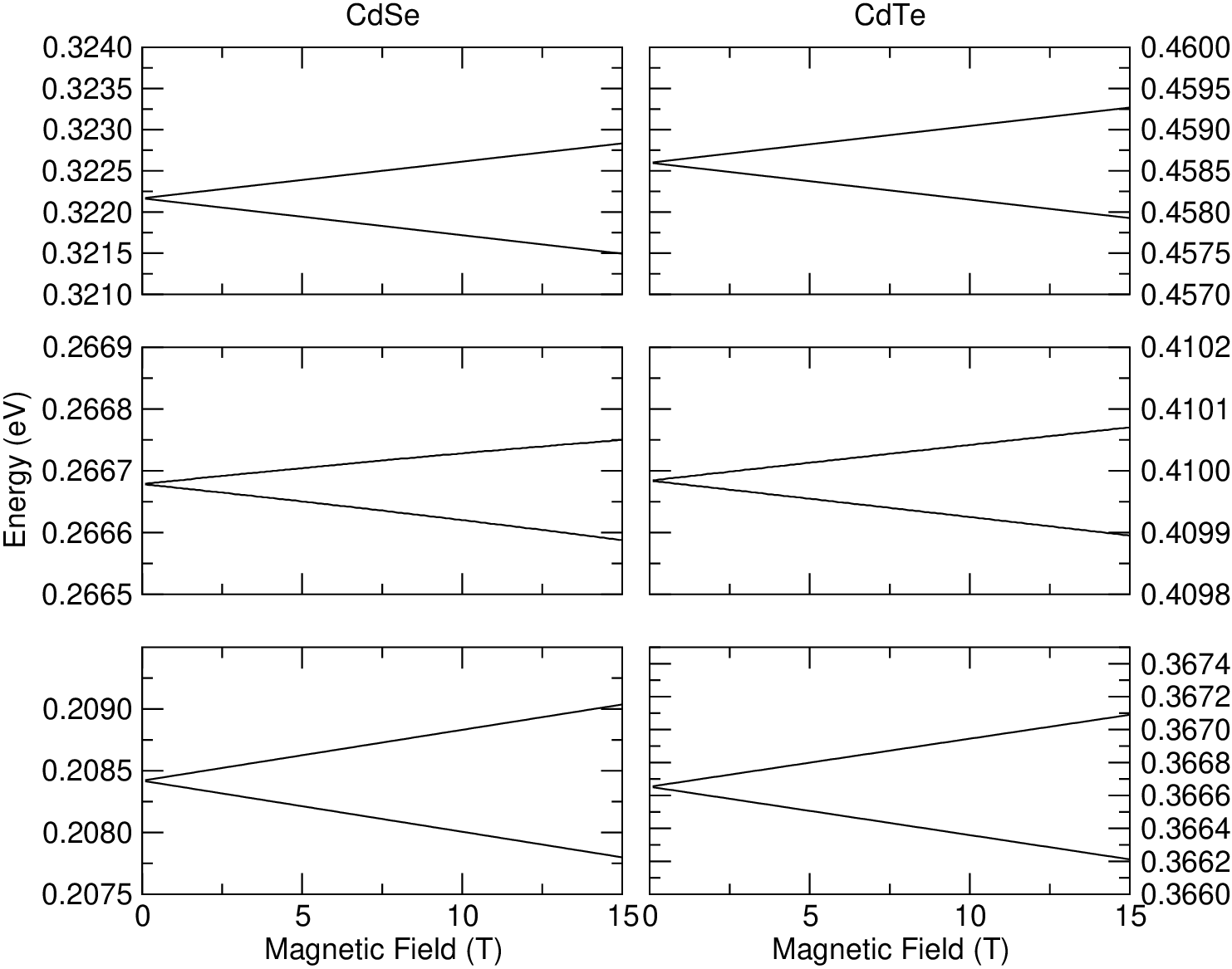}
  \caption{Hole Zeeman splitting for 777 atoms CdSe and CdTe nanostructures
    \label{Fig:Zeeman777h}}
\end{figure}

\begin{figure}
  \includegraphics[scale=0.30,angle=0]{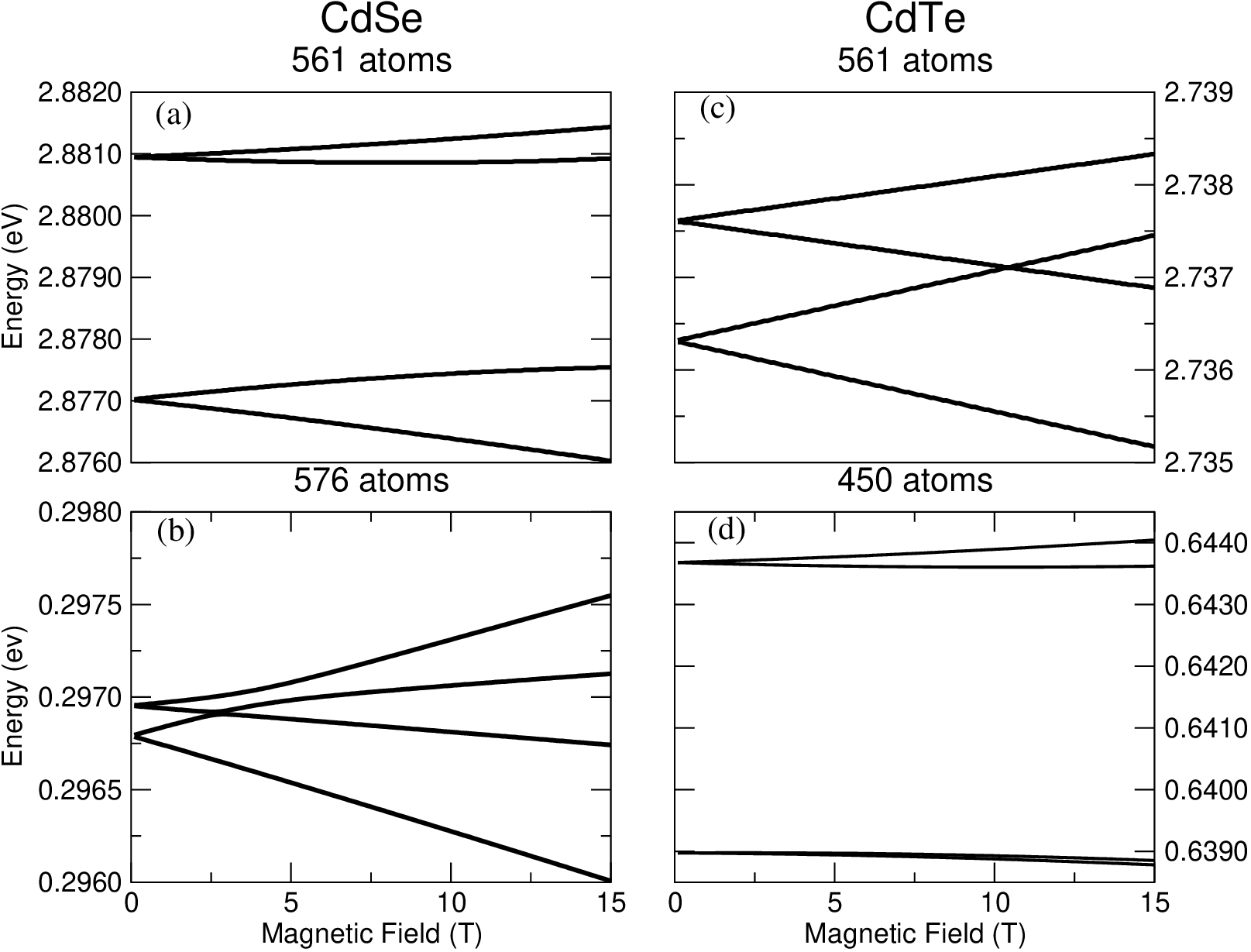}
  \caption{Non-linear Zeeman splittings. 
  (a) Electrons in 561 atoms CdSe nanostructure. 
  (b) Holes in 576 atoms CdSe nanostructure. 
  (c) Electrons in 561 atoms CdTe nanostructure. 
  (d) Holes in 450 atoms CdTe nanostructure. 
  \label{Fig:Zeeman_nonlinear}}
\end{figure}

For zero field energy levels, the band edge states show redshifting as
the number of atoms in nanostructures increases. The valence band edge
state occurs as heavy-hole like state for some of the nanostructures
while occurs as light-hole like state for other nanostructures, even
when there is only minor difference in the numbers of atoms contained
in nanostructures. This result indicates that the shape of the
nanostructure has profound impact on the spin content of the hole
states.  No degeneracy except for the Kramers' degeneracy is observed,
which we attribute to the strong quantum confinement effect of the
small nanostructures.  In Fig.\ref{Fig:Zeeman777}
(Fig.\ref{Fig:Zeeman777h}) we plot the magnetic field dependence of
several electron (hole) energy levels for 777 atoms CdSe and CdTe
nanostructures.  The magnetic field is pointing along the z-direction,
which coincides with the c-axis of the wurtzite structure.  As seen in
these figures, the Zeeman splitting is nearly linear in B.  Magnetic
field dependence for most of the electron and hole levels calculated
in this work show qualitatively the same behavior, while
quantitatively the slope of Zeeman splitting varies, giving rise to
different g-factors.  However some of the electron and hole levels
show strongly non-linear dispersions. In Fig.
\ref{Fig:Zeeman_nonlinear} we plot the Zeeman splitting for some of
those levels. The non-linear dispersions occur when the intrinsic
Zeeman spltting is close to the inter-level spacing, giving rise to
level anti-crossing, or when the intrinsic Zeeman splitting is small
compared to the splitting induced by inter-doublets couplings.

\begin{figure}
  \includegraphics[scale=0.35,angle=0]{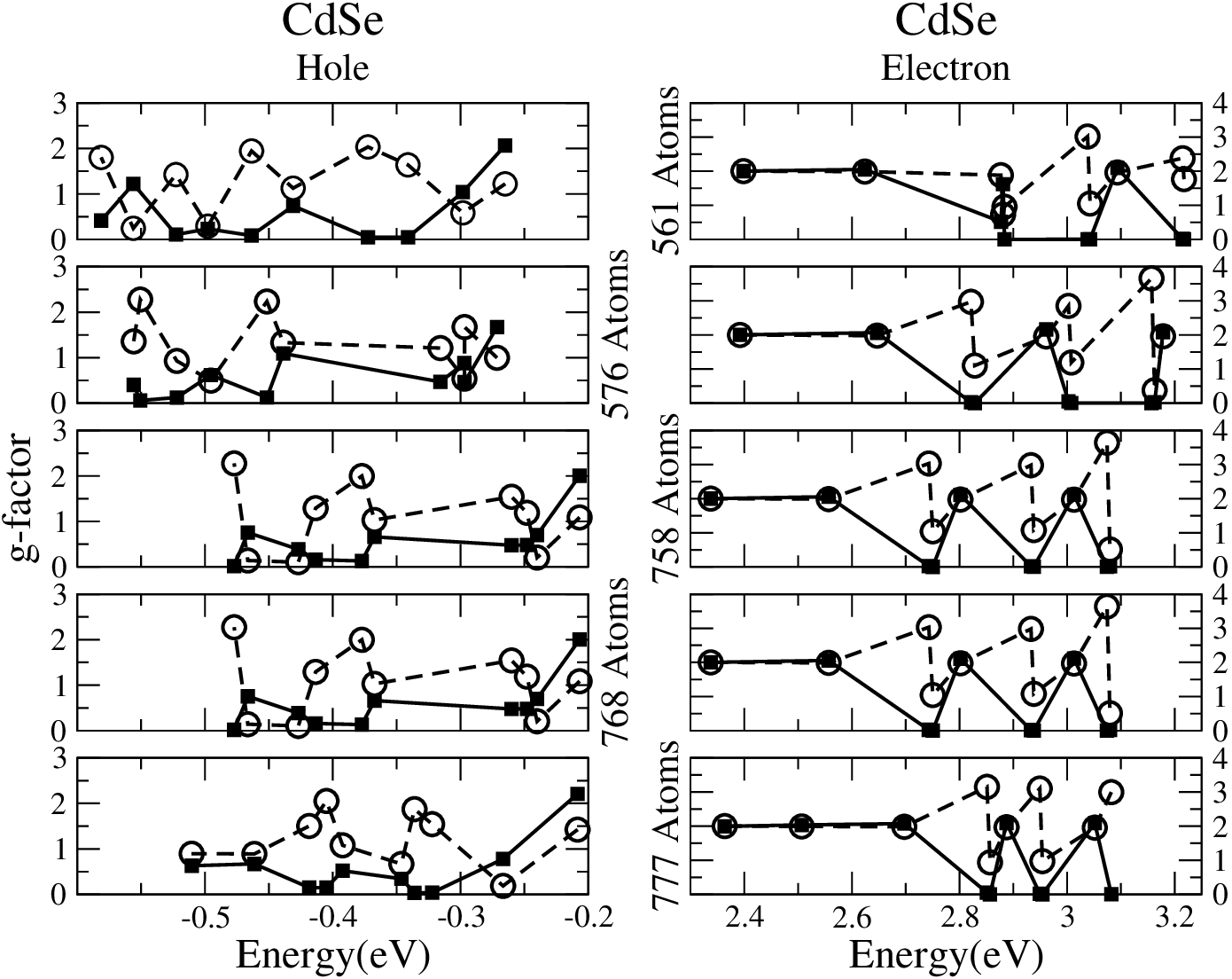}
  \caption{Intrinsic g-factor of first ten electron and hole states 
  for CdSe nanostructures at various sizes. 
  Filled circles represent $g_z$ and empty circles represent $g_x$.
  \label{Fig:g-factorCdSe}}
\end{figure}

\begin{figure}
  \includegraphics[scale=0.35,angle=0]{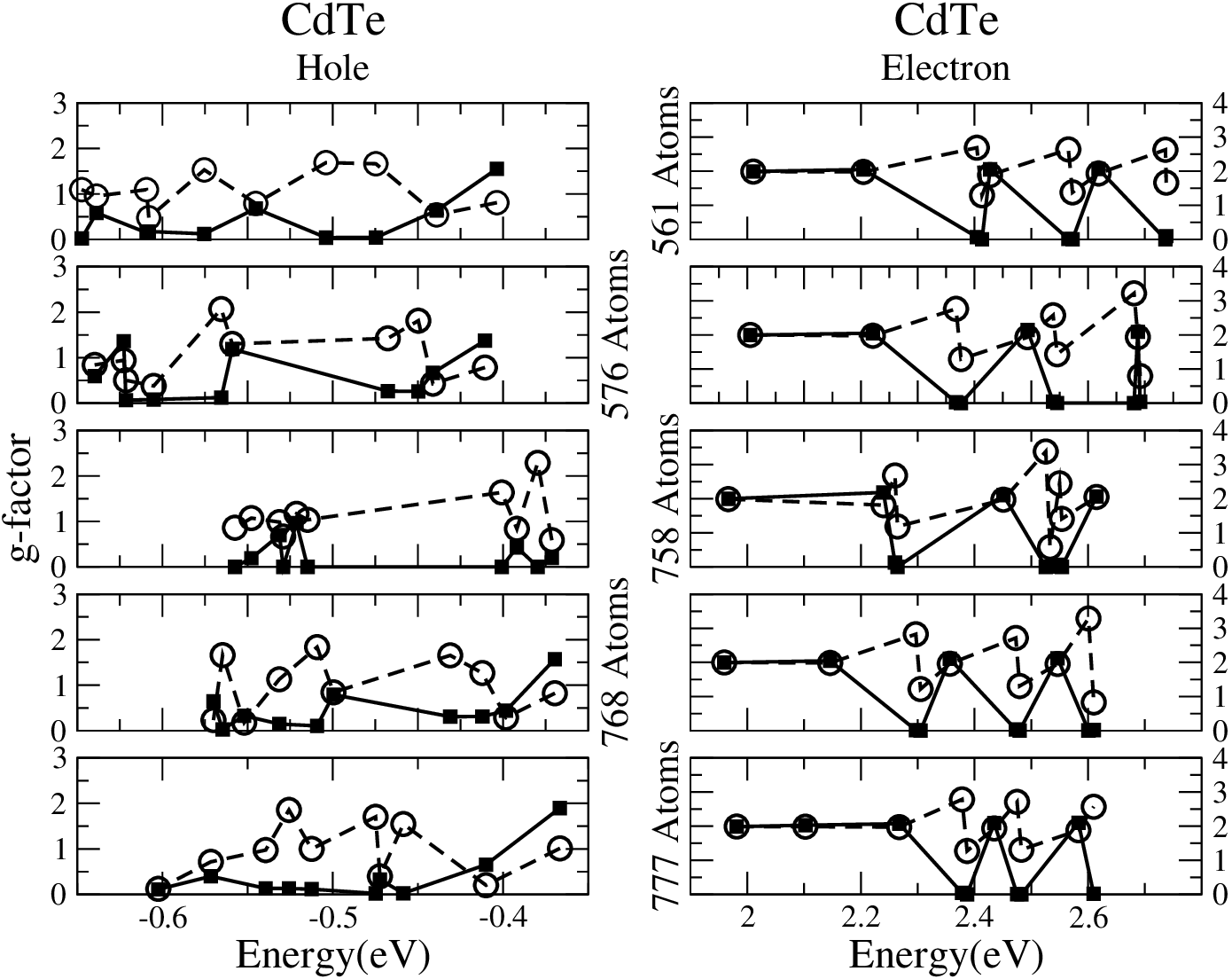}
  \caption{Intrinsic g-factor of first ten electron and hole states
  for CdTe nanostructures at various sizes. 
  Filled circles represent $g_z$ and empty circles represent $g_x$.
  \label{Fig:g-factorCdTe}}
\end{figure}

In Fig.\ref{Fig:g-factorCdSe} (Fig.\ref{Fig:g-factorCdTe}) we plot the
intrinsic g-factor of first ten electron and hole states for CdSe
(CdTe) nanostructures at various sizes. We found that $g_x\approx g_y
\neq g_z$ and $g_y$ is not plotted for clarity.  In Table
\ref{Table:size} we summarize the size and aspect ratio of the
nanostructures used in Fig.\ref{Fig:g-factorCdSe} and
Fig.\ref{Fig:g-factorCdTe}. The aspect ratio is defined to be the
ratio between $L_z$ and the average diameter $\sqrt{L_x L_y}$ in x-y
plane.  The prominent features are the redshift (blueshift) of
electron (hole) levels and anisotropy of g-factors.  Quantitatively we
found it difficult to identify a simple scaling relation between the
g-factor and size or aspect ratio of nanostructures using the data
shown here and other data for smaller size nanostructures. The size
($L_x, L_y, L_z$) and the aspect ratio are average quantities which
are used to characterize the nanostructures.  For higher dimensional
systems, these average quantities have been successfully used to
characterize the g-factors, since the detail of the boundary is less
significant. However for small size nanostructures the detail of the
structure is important in determining the energy levels,
wavefunctions, and hence the g-factors. For example, it is difficult
to modify the aspect ratio along without altering substantially the
detail shape of the structure. We speculate that this is the main
reason to make it difficult to identify a simple scaling relation
between g-factor and other averaged quantities.  It is thus
advantageous to use tight-binding method, which captures the detail of
the structure, instead of effective mass approximation type methods.


Qualitatively some observations can be made. First since
the g-factor is not a smooth function of the eigenenergy,
nearby states might possess very different g-factors. 
The g-factors also change substantially as the size changes. 
We observe that some energy levels move toward higher energy 
while some levels move toward lower energy as one increases
the aspect ratio. Similar dependence has been recently observed
in the calculation of electronic states of CdSe quantum rods.
\cite{jiangtaohu99}. We also observe that the electron g-factors
are less sensitive than the hole g-factors to the
aspect ratio. This can be attributed to the fact that the $S=\frac{1}{2}$
like behavior of the electron is insensitive to the aspect ratio 
while for the $J=\frac{1}{2}$ like holes, the mixing of heavy-hole
and light-hole is sensitive to the aspect ratio.


\begin{table}
  \caption{\label{Table:size} Size and aspect ratio of the nanostructures.}
  \begin{tabular}{cccccc}
    \hline \hline
    Number of atoms           
    &450   &561   &758   &768   &777   \\ 
    \hline
    $\sqrt{L_x L_y}$ (\AA)    
    &26.76 &22.85 &34.55 &27.42 &20.44 \\ 
    $L_z$  (\AA)              
    &21.88 &35.88 &35.88 &39.38 &49.88 \\ 
    Aspect Ratio              
    &0.82  &1.64  &1.04  &1.44  &2.44  \\ 
    \hline \hline
  \end{tabular}
\end{table}

\section{Summary and Discussion}
In summary we have developed a general method to evaluate electron and
hole Zeeman splittings in semiconductor nanostructures within a 
tight-binding framework. The calculation is carried out
within the electron-hole picture instead of the conduction-valence band 
electron picture.  Hence the scheme can be readily extended to
the exciton Zeeman splitting calculation by including the
electron-hole Coulomb interaction.  The results found here are qualitatively
similar to the results in Ref \onlinecite{chen:045311}, in which only
band edge states are calculated. The quantitative difference may be
attributed to the approximation used during the transformation to the
electron-hole picture. In most cases we observe nearly linear Zeeman 
splittings induced by the external magnetic field. 
In some cases we also observe
non-linear Zeeman splittings which are due to the interlevel coupling
between nearby Kramers' doublets. Those results are qualitatively similar to what
are obtained in Ref \onlinecite{prado:201310}.

We find that the behavior of electron Zeeman splittings of nanostructures 
with different aspect ratio does not show drastic difference.
This is in contrast to the drastic difference (change of sign) between 
spherical quantum dot (SQD) and
semi-spherical quantum dot (SSQD) found in Ref \onlinecite{prado:201310}.
It should be
noted that in Ref \onlinecite{prado:201310}, SSQD is treated by
imposing boundary condition on envelope function, requiring envelope
function to be vanishing in the upper hemisphere.  
This boundary condition is very different from the boundary condition
used here for nanostructures with different aspect ratio.
In other word, the idea of SSQD in Ref  \onlinecite{prado:201310} is different from
the idea of quasi-spherical regime for the elongated nanostructure with 
varying aspect ratio in other references.\cite{schrier:235301,rodina:155312,chen:045311}

There may be strong difference between magneto-optical properties of
free electron-hole pairs and excitons, in which the Coulomb interaction
is included.
Two key questions to be addressed are the effect of the Coulomb
interaction and how to experimentally observe electron, hole, and exciton 
Zeeman splittings using magneto-absorption spectroscopy.
To answer the first question one has to estimate how strongly the electron-hole
Coulomb interaction modifies the free electron-hole pair picture.
In Ref \onlinecite{leung98}, a restricted basis, single excitation 
configuration interaction
representation method is used to calculate the exciton fine structure for CdSe
nanostructures. It was shown that about 18 electron and 24 hole single-particle
states needed to be included to get a reasonable accurate exciton energy.
We expect that roughly the same number of states will be needed in order to 
accurately evaluate the exciton Zeeman splittings. Since our results indicate that 
the g-factor of electrons and holes may vary substantially from one
state to another, we expect that the exciton g-factor may be very different
from the simple sum of the g-factor of free electrons and holes.
Starting from the general framework developed here, the effect of
Coulomb interaction can be included via configuration interaction method\cite{leung97,leung98}
or time-dependent tight-binding method.\cite{Hill96}

To shed some light on the second question we have to consider the
problem of optical orientation in nanostructures.
In bulk or quantum well materials, one typically assumes that the conduction 
band edge is s-like while the valence band edge is heavy-hole like.
In this case the optically active excitons only consist of heavy-holes.
Due to the angular momentum conservation, the optically excited 
electron and heavy-hole spin will be aligned in particular configuration depending
on the polarization of the optical pulse.
However, in nanostructures the strong heavy-hole light-hole
mixing and the Coulomb interaction break down this naive picture,
as it becomes difficult to determine the spin configuration of the constituent
electrons and holes of the excitons.
A realistic calculation of the optical absorption oscillator strength might be needed
in order to fully understand what is actually measured in the experiment.

\begin{acknowledgments}
  We thank Joshua Schrier for his critical reading of the manuscript. 
  We acknowledge the support of National Science Council in Taiwan through
  grant NSC 93-2112-M-007-038.
\end{acknowledgments}


\begin{thebibliography}{17}
\expandafter\ifx\csname natexlab\endcsname\relax\def\natexlab#1{#1}\fi
\expandafter\ifx\csname bibnamefont\endcsname\relax
  \def\bibnamefont#1{#1}\fi
\expandafter\ifx\csname bibfnamefont\endcsname\relax
  \def\bibfnamefont#1{#1}\fi
\expandafter\ifx\csname citenamefont\endcsname\relax
  \def\citenamefont#1{#1}\fi
\expandafter\ifx\csname url\endcsname\relax
  \def\url#1{\texttt{#1}}\fi
\expandafter\ifx\csname urlprefix\endcsname\relax\def\urlprefix{URL }\fi
\providecommand{\bibinfo}[2]{#2}
\providecommand{\eprint}[2][]{\url{#2}}

\bibitem[{\citenamefont{Zutic et~al.}(2004)\citenamefont{Zutic, Fabian, and
  Sarma}}]{zutic:323}
\bibinfo{author}{\bibfnamefont{I.}~\bibnamefont{Zutic}},
  \bibinfo{author}{\bibfnamefont{J.}~\bibnamefont{Fabian}}, \bibnamefont{and}
  \bibinfo{author}{\bibfnamefont{S.~D.} \bibnamefont{Sarma}},
  \bibinfo{journal}{Rev. Mod. Phys.} \textbf{\bibinfo{volume}{76}},
  \bibinfo{eid}{323} (\bibinfo{year}{2004}).

\bibitem[{\citenamefont{{Michael A. Nielsen} and {Isaac L. Chuang
  }}(2000)}]{Nielsen00}
\bibinfo{author}{\bibnamefont{{Michael A. Nielsen}}} \bibnamefont{and}
  \bibinfo{author}{\bibnamefont{{Isaac L. Chuang }}},
  \emph{\bibinfo{title}{Quantum Computation and Quantum Information}}
  (\bibinfo{publisher}{Cambridge University Press}, \bibinfo{year}{2000}).

\bibitem[{\citenamefont{Kosaka et~al.}(2001)\citenamefont{Kosaka, Kiselev,
  Baron, Kim, and Yablonovitch}}]{kosaka2001}
\bibinfo{author}{\bibfnamefont{H.}~\bibnamefont{Kosaka}},
  \bibinfo{author}{\bibfnamefont{A.~A.} \bibnamefont{Kiselev}},
  \bibinfo{author}{\bibfnamefont{F.~A.} \bibnamefont{Baron}},
  \bibinfo{author}{\bibfnamefont{K.~W.} \bibnamefont{Kim}}, \bibnamefont{and}
  \bibinfo{author}{\bibfnamefont{E.}~\bibnamefont{Yablonovitch}},
  \bibinfo{journal}{Electronics Letters} \textbf{\bibinfo{volume}{37}},
  \bibinfo{pages}{464} (\bibinfo{year}{2001}).

\bibitem[{\citenamefont{Peng et~al.}(2000)\citenamefont{Peng, Manna, Yang,
  Wickham, Scher, Kadavanich, and Alivisatos}}]{xiaogangpeng00}
\bibinfo{author}{\bibfnamefont{X.}~\bibnamefont{Peng}},
  \bibinfo{author}{\bibfnamefont{L.}~\bibnamefont{Manna}},
  \bibinfo{author}{\bibfnamefont{W.}~\bibnamefont{Yang}},
  \bibinfo{author}{\bibfnamefont{J.}~\bibnamefont{Wickham}},
  \bibinfo{author}{\bibfnamefont{E.}~\bibnamefont{Scher}},
  \bibinfo{author}{\bibfnamefont{A.}~\bibnamefont{Kadavanich}},
  \bibnamefont{and} \bibinfo{author}{\bibfnamefont{A.~P.}
  \bibnamefont{Alivisatos}}, \bibinfo{journal}{Nature}
  \textbf{\bibinfo{volume}{404}}, \bibinfo{pages}{59} (\bibinfo{year}{2000}).

\bibitem[{\citenamefont{Manna et~al.}(2000)\citenamefont{Manna, Scher, and
  Alivisatos}}]{manna00}
\bibinfo{author}{\bibfnamefont{L.}~\bibnamefont{Manna}},
  \bibinfo{author}{\bibfnamefont{E.~C.} \bibnamefont{Scher}}, \bibnamefont{and}
  \bibinfo{author}{\bibfnamefont{A.~P.} \bibnamefont{Alivisatos}},
  \bibinfo{journal}{J. Am. Chem. Soc.} \textbf{\bibinfo{volume}{122}},
  \bibinfo{pages}{12700} (\bibinfo{year}{2000}).

\bibitem[{\citenamefont{Kotlyar et~al.}(2001)\citenamefont{Kotlyar, Reinecke,
  Bayer, and Forchel}}]{kotlyar:085310}
\bibinfo{author}{\bibfnamefont{R.}~\bibnamefont{Kotlyar}},
  \bibinfo{author}{\bibfnamefont{T.~L.} \bibnamefont{Reinecke}},
  \bibinfo{author}{\bibfnamefont{M.}~\bibnamefont{Bayer}}, \bibnamefont{and}
  \bibinfo{author}{\bibfnamefont{A.}~\bibnamefont{Forchel}},
  \bibinfo{journal}{Phys. Rev. B} \textbf{\bibinfo{volume}{63}},
  \bibinfo{eid}{085310} (\bibinfo{year}{2001}).

\bibitem[{\citenamefont{Prado et~al.}(2004)\citenamefont{Prado, Trallero-Giner,
  Alcalde, Lopez-Richard, and Marques}}]{prado:201310}
\bibinfo{author}{\bibfnamefont{S.~J.} \bibnamefont{Prado}},
  \bibinfo{author}{\bibfnamefont{C.}~\bibnamefont{Trallero-Giner}},
  \bibinfo{author}{\bibfnamefont{A.~M.} \bibnamefont{Alcalde}},
  \bibinfo{author}{\bibfnamefont{V.}~\bibnamefont{Lopez-Richard}},
  \bibnamefont{and} \bibinfo{author}{\bibfnamefont{G.~E.}
  \bibnamefont{Marques}}, \bibinfo{journal}{Phys. Rev. B}
  \textbf{\bibinfo{volume}{69}}, \bibinfo{eid}{201310} (\bibinfo{year}{2004}).

\bibitem[{\citenamefont{Rodina et~al.}(2003)\citenamefont{Rodina, Efros, and
  Alekseev}}]{rodina:155312}
\bibinfo{author}{\bibfnamefont{A.~V.} \bibnamefont{Rodina}},
  \bibinfo{author}{\bibfnamefont{A.~L.} \bibnamefont{Efros}}, \bibnamefont{and}
  \bibinfo{author}{\bibfnamefont{A.~Y.} \bibnamefont{Alekseev}},
  \bibinfo{journal}{Phys. Rev. B} \textbf{\bibinfo{volume}{67}},
  \bibinfo{eid}{155312} (\bibinfo{year}{2003}).

\bibitem[{\citenamefont{Chen and Whaley}(2004)}]{chen:045311}
\bibinfo{author}{\bibfnamefont{P.}~\bibnamefont{Chen}} \bibnamefont{and}
  \bibinfo{author}{\bibfnamefont{K.~B.} \bibnamefont{Whaley}},
  \bibinfo{journal}{Phys. Rev. B} \textbf{\bibinfo{volume}{70}},
  \bibinfo{eid}{045311} (\bibinfo{year}{2004}).

\bibitem[{\citenamefont{Schrier and Whaley}(2003)}]{schrier:235301}
\bibinfo{author}{\bibfnamefont{J.}~\bibnamefont{Schrier}} \bibnamefont{and}
  \bibinfo{author}{\bibfnamefont{K.~B.} \bibnamefont{Whaley}},
  \bibinfo{journal}{Phys. Rev. B} \textbf{\bibinfo{volume}{67}},
  \bibinfo{eid}{235301} (\bibinfo{year}{2003}).

\bibitem[{\citenamefont{Gupta et~al.}(1999)\citenamefont{Gupta, Awschalom,
  Peng, and Alivisatos}}]{gupta98}
\bibinfo{author}{\bibfnamefont{J.~A.} \bibnamefont{Gupta}},
  \bibinfo{author}{\bibfnamefont{D.~D.} \bibnamefont{Awschalom}},
  \bibinfo{author}{\bibfnamefont{X.}~\bibnamefont{Peng}}, \bibnamefont{and}
  \bibinfo{author}{\bibfnamefont{A.~P.} \bibnamefont{Alivisatos}},
  \bibinfo{journal}{Phys. Rev. B} \textbf{\bibinfo{volume}{59}},
  \bibinfo{pages}{R10421} (\bibinfo{year}{1999}).

\bibitem[{\citenamefont{{H. Haken}}(1983)}]{Haken83}
\bibinfo{author}{\bibnamefont{{H. Haken}}}, \emph{\bibinfo{title}{Quantum
  {F}ield {T}heory of {S}olids}} (\bibinfo{publisher}{North-Holland,
  Amsterdam}, \bibinfo{year}{1983}).

\bibitem[{\citenamefont{Leung et~al.}(1998)\citenamefont{Leung, Pokrant, and
  Whaley}}]{leung98}
\bibinfo{author}{\bibfnamefont{K.}~\bibnamefont{Leung}},
  \bibinfo{author}{\bibfnamefont{S.}~\bibnamefont{Pokrant}}, \bibnamefont{and}
  \bibinfo{author}{\bibfnamefont{K.~B.} \bibnamefont{Whaley}},
  \bibinfo{journal}{Phys. Rev. B} \textbf{\bibinfo{volume}{57}},
  \bibinfo{pages}{12291} (\bibinfo{year}{1998}).

\bibitem[{\citenamefont{Boykin et~al.}(2001)\citenamefont{Boykin, Bowen, and
  Klimeck}}]{boykin01}
\bibinfo{author}{\bibfnamefont{T.~B.} \bibnamefont{Boykin}},
  \bibinfo{author}{\bibfnamefont{R.~C.} \bibnamefont{Bowen}}, \bibnamefont{and}
  \bibinfo{author}{\bibfnamefont{G.}~\bibnamefont{Klimeck}},
  \bibinfo{journal}{Phys. Rev. B} \textbf{\bibinfo{volume}{63}},
  \bibinfo{eid}{245314} (\bibinfo{year}{2001}).

\bibitem[{\citenamefont{Hu et~al.}(2002)\citenamefont{Hu, Wang, Li, Yang, and
  Alivisatos}}]{jiangtaohu99}
\bibinfo{author}{\bibfnamefont{J.}~\bibnamefont{Hu}},
  \bibinfo{author}{\bibfnamefont{L.-W.} \bibnamefont{Wang}},
  \bibinfo{author}{\bibfnamefont{L.-S.} \bibnamefont{Li}},
  \bibinfo{author}{\bibfnamefont{W.}~\bibnamefont{Yang}}, \bibnamefont{and}
  \bibinfo{author}{\bibfnamefont{A.~P.} \bibnamefont{Alivisatos}},
  \bibinfo{journal}{J. Phys. Chem. B} \textbf{\bibinfo{volume}{106}},
  \bibinfo{pages}{2447} (\bibinfo{year}{2002}).

\bibitem[{\citenamefont{Leung and Whaley}(1997)}]{leung97}
\bibinfo{author}{\bibfnamefont{K.}~\bibnamefont{Leung}} \bibnamefont{and}
  \bibinfo{author}{\bibfnamefont{K.~B.} \bibnamefont{Whaley}},
  \bibinfo{journal}{Phys. Rev. B} \textbf{\bibinfo{volume}{56}},
  \bibinfo{pages}{7455} (\bibinfo{year}{1997}).

\bibitem[{\citenamefont{Hill and Whaley}(1996)}]{Hill96}
\bibinfo{author}{\bibfnamefont{N.~A.} \bibnamefont{Hill}} \bibnamefont{and}
  \bibinfo{author}{\bibfnamefont{K.~B.} \bibnamefont{Whaley}},
  \bibinfo{journal}{Chem. Phys.} \textbf{\bibinfo{volume}{210}},
  \bibinfo{pages}{117} (\bibinfo{year}{1996}).

\end{thebibliography}

\end{document}